\documentclass[prl,floatfix,twocolumn,preprintnumbers,amsmath,amssymb,superscriptaddress]{revtex4}
\usepackage{color}

\usepackage{graphicx,psfrag}
\usepackage{hyperref}
\usepackage{verbatim}
\usepackage[utf8]{inputenc}
\newcommand{\ti}{\times}
\newcommand{\mc}{\mathcal}

\begin{document}
\title{Consistency of Hitomi, XMM-Newton and Chandra 3.5 keV data from Perseus}
\author{Joseph P. Conlon}
\email{joseph.conlon@physics.ox.ac.uk}
\affiliation{Rudolf Peierls Centre for Theoretical Physics, University of Oxford,
1 Keble Rd., Oxford OX1 3NP, UK}
\author{Francesca Day}
\email{francesca.day@physics.ox.ac.uk}
\affiliation{Rudolf Peierls Centre for Theoretical Physics, University of Oxford,
1 Keble Rd., Oxford OX1 3NP, UK}
\author{Nicholas Jennings}
\email{nicholas.jennings@physics.ox.ac.uk}
\affiliation{Rudolf Peierls Centre for Theoretical Physics, University of Oxford,
1 Keble Rd., Oxford OX1 3NP, UK}
\author{Sven Krippendorf}
\email{sven.krippendorf@physics.ox.ac.uk}
\affiliation{Rudolf Peierls Centre for Theoretical Physics, University of Oxford,
1 Keble Rd., Oxford OX1 3NP, UK}
\author{Markus Rummel}
\email{rummelm@mcmaster.ca}
\affiliation{Rudolf Peierls Centre for Theoretical Physics, University of Oxford,
1 Keble Rd., Oxford OX1 3NP, UK}
\begin{abstract}

\emph{Hitomi} observations of Perseus with the Soft X-ray Spectrometer (SXS) provide a high-resolution look at the 3.5 keV feature
reported by multiple groups in the Perseus cluster. The \emph{Hitomi} spectrum -- which involves the sum of diffuse cluster emission
and the point-like central Active Galactic Nucleus (AGN) -- does not show any excess at $E \sim 3.5 {\rm keV}$, giving an apparent inconsistency
with previous observations of excess diffuse emission. We point out that 2009 \emph{Chandra} data reveals a strong
dip in the AGN spectrum at $E  = (3.54 \pm 0.02) {\rm keV}$ (cluster frame) -- the identical energy to the diffuse excess observed by \emph{XMM-Newton}.
Scaling this dip to the 2016 AGN luminosity and adding it to
the diffuse \emph{XMM-Newton} excess, this predicts an overall dip in the SXS field of view  of
$(-5.9 \pm 4.4) \ti 10^{-6} \, {\rm ph} \, {\rm cm}^{-2} \, {\rm s}^{-1}$ at $E= 3.54$ keV -- a precise match to
the \emph{Hitomi} data when broadened by the dark matter virial velocity.
We describe models of Fluorescent Dark Matter that can reproduce this physics, in which dark matter absorbs and then re-emits 3.5 keV photons
emitted from the central AGN.

\end{abstract}
\maketitle

The nature of dark matter is a question of fundamental importance in particle physics, astrophysics and cosmology.
The X-ray spectra  around $E = 3.5 \, {\rm keV}$ of galaxy clusters and other objects have been intensely studied following
the announcement in \cite{Bulbul:2014sua, Boyarsky:2014jta} of an unidentified emission line around this energy.
In this respect, the Perseus cluster has been of special interest. For both \emph{XMM-Newton} and \emph{Chandra} data,
the signal for Perseus found in \cite{Bulbul:2014sua, Boyarsky:2014jta}
was much stronger than that inferred from the stacked sample of distant clusters in \cite{Bulbul:2014sua} (in the much studied scenario where the 3.5 keV line arises from dark matter decay to photons).
A similar result was found in \cite{1411.1758}, and also confirmed with
\emph{Suzaku} studies of the Perseus cluster in \cite{Urban:2014yda, Franse:2016dln} (although see
\cite{Tamura:2014mta} for a contrary view on \emph{Suzaku} data).

The 3.5 keV line has been discussed at length in the literature. Both astrophysical explanations (such as potassium emission \cite{Bulbul:2014sua, 1411.1758} or sulphur charge exchange \cite{GuKaastra}) and more exotic ones such as dark matter have been put forward as its origin.
There is currently no consensus opinion on the physics behind the line and
an overall review of studies of the 3.5 keV line is contained in \cite{151000358}.

The CCD technology present on \emph{Chandra}, \emph{XMM-Newton} and \emph{Suzaku} has limited energy resolution.
The launch of \emph{Hitomi}, with the microcalorimeters present on its Soft X-ray Spectrometer (SXS), was expected
to offer a decisive test of the 3.5 keV line, both in Perseus and in other clusters. Sadly \emph{Hitomi} was lost around one month after launch,
but prior to its loss it performed a deep groundbreaking observation of the centre of the Perseus cluster.

The data from the \emph{Hitomi} observations around $E \sim 3.5 \, {\rm keV}$ were recently reported in \cite{Aharonian:2016gzq}.
No excess was seen, and it is stated that this is incompatible at more than 3$\sigma$  with the
strong signal from Perseus observed with other telescopes.

In this paper, we show that that there is no incompatibility and that the
overall observational picture is consistent.
While the \emph{XMM-Newton} and \emph{Chandra} spectra involve only diffuse cluster emission,
the \emph{Hitomi} spectrum contains the sum of diffuse and AGN emission.
2009 \emph{Chandra} data reveals a dip in the AGN spectrum around 3.5 keV~\cite{Berg:2016ese}.
Here we show that this dip is well described by an absorption feature located at $(3.54 \pm 0.02)$ keV (we quote all energies in the cluster frame).
In this case, the combination of this dip together with the previously observed
diffuse excess results overall in a mild dip at 3.54 keV, as observed in the \emph{Hitomi} spectrum.

\subsection{Observational Sensitivity}

The micro-calorimeters on \emph{Hitomi} provide unprecedented energy resolution, $\Delta E \sim 5 \, {\rm eV}$, and
also have a rapid readout time of $\mathcal{O}(10^{-5})$s ensuring the absence of
pileup. Pileup occurs when the rate of incident photons is greater than the maximum readout rate. This can lead to the energies of multiple photons being summed and recorded as a single photon, as well as other undesirable effects. Pileup results in loss of information about the source spectrum, and highly piled up spectra are unusable in searches for spectral features. However, \emph{Hitomi}'s angular resolution
is poor compared to \emph{Chandra} and \emph{XMM-Newton}, with a half-power diameter of $1.2'$.
The SXS observations of the Perseus cluster \cite{Aharonian:2016pyf, Aharonian:2016gzq}
covered an (almost) $3' \ti 3'$ region at the centre of the Perseus cluster, including the central cluster AGN at the core of the galaxy NGC1275. It is reported in
\cite{Aharonian:2016gzq} that the AGN contributes 15\% of the total counts to the 3-4 keV
spectrum (the poor angular resolution of \emph{Hitomi} makes it not possible to isolate and remove
the AGN as a point source, as can be done for \emph{Chandra} and \emph{XMM-Newton} data).
As a result, the \emph{Hitomi} data reported in \cite{Aharonian:2016gzq} is sensitive to the sum of any 3.5 keV features in the diffuse
emission \emph{plus} any 3.5 keV features in the AGN spectrum.

In contrast, \emph{XMM-Newton} and \emph{Chandra} both use CCD technology with $\Delta E \sim 100 \, {\rm eV}$, but have
far superior optics to \emph{Hitomi} (half-power diameters of $17''$ and
$1''$ respectively). In this case, it is possible to analyse just the diffuse
 emission, as the central AGN can be isolated and subtracted
as a point source (through e.g.~\emph{wavdetect}).

In general, there are complications in studying the AGN spectrum. For both \emph{XMM-Newton} observations of the Perseus cluster (taken in 2001 and 2006) and for all ACIS-S
\emph{Chandra} observations taken in 2002-4, the AGN was on-axis. This causes significant pileup, distorting the photon spectrum.
Of these cases, pileup is smallest for the relatively short 2001 \emph{XMM-Newton} observation. However, here
the AGN was at its weakest luminosity in observational history, giving a relatively poor contrast against the cluster emission.

The only \emph{Chandra} observations without significant pileup in the AGN spectrum are a crucial set of four observations taken in 2009 and totalling 200ks
(\emph{Chandra} obsids 11713, 12025, 12033, 12036). For these observations the AGN is
located at the edge of the ACIS-I field of view, approximately seven arcminutes
away from the optical axis. The optical distortions -- a consequence of the large separation from the focal point -- spread the AGN image over an ellipse of approximate radii $11''$ and $7''$; the small \emph{Chandra} pixel size
of $0.5'' \ti 0.5''$ then ensures that pileup is minimal for these observations, while the brightness of the AGN ensures a high contrast against the background
cluster emission.

\subsection{Data Analysis}

The data for these observations is publicly available from the \emph{Chandra} X-ray archive. We download it and reprocess it using the calibration software CALDB 4.6.9, and analyse it using the standard \emph{Chandra} software packages
CIAO 4.8 (for data reduction and processing) and Sherpa (for data analysis and fitting of models) \cite{ciao, sherpa}. For the relevant observations (\emph{Chandra} obsids 11713, 12025, 12033, 12036) the AGN is visible as a very bright point source
close to the edge of the chip (although far enough away that it does not dither off the edge).
The observations are unaffected by flares and we use the whole of the available time.

We extract the spectra from an ellipse around NGC1275 of radii $11.6''$ and $7.2''$, stacking the four observations together.
The resulting spectrum is the sum of the intrinsic emission of the AGN plus cluster emission from within the extraction region.

One can subtract the cluster emission using the spectrum from a region close to the AGN (this was the approach used in \cite{Berg:2016ese} and yields very similar results to those presented here).
However, it is statistically preferable to model the cluster emission directly,
rather than trying to remove it via subtraction.

The recent \emph{Hitomi} observations have provided a highly accurate characterisation of the thermal emission in the core of the Perseus cluster.
In particular, as determined in \cite{Aharonian:2016gzq} the \emph{Hitomi} spectrum for the thermal emission is well fit by a single-component \emph{bapec} thermal plasma with a temperature of $kT = (3.48 \pm 0.07) {\rm keV}$, an abundance of $Z = (0.54 \pm 0.03)$ and a velocity dispersion of $179 \pm 16 {\rm km} \, {\rm s}^{-1}$. \emph{bapec} refers to the photon emission spectrum produced by thermal bremsstrahlung from an ionised plasma at a temperature $T$ with a metallicity $Z$, broadened by a given velocity dispersion, applicable to the emission from the hot gas within galaxy clusters.

The AGN spectrum is modelled as a power law, and both the AGN and thermal emission is absorbed by neutral hydrogen in the Milky Way. Our approach is therefore to use Sherpa to fit our extracted spectrum with the (absorbed) sum of power-law and thermal emission, \emph{xswabs} $\ti $ (\emph{powlaw1d + xsbapec}). These refer to the implementation of various standard physical models within Sherpa. Physically, \emph{xswabs} refers to photoelectric absorption of X-rays within the Milky Way,
as measured by the column density of Neutral Hydrogen,
$\exp (- n_H \sigma(E))$, $\emph{powlaw1d}$ refers to a power-law distribution of photons, $N_{\gamma}(E) \propto E^{-\gamma}$, and 
$\emph{xsbapec}$ refers to the implementation of the $\emph{bapec}$ model for velocity-broadened thermal bremsstrahlung emission from the ionised
plasma of the intracluster medium (which does not have a simple analytic form).
We use the \emph{chi2datavar} statistic of the Sherpa fitting package (i.e. determining the variances from the data itself). We
 freeze the parameters of the \emph{bapec} model to be those determined by the \emph{Hitomi} satellite\cite{Aharonian:2016gzq}, allowing only the amplitude to float.  Given the very high brightness of the source, there is negligible contamination from any of the local hot bubble, the diffuse cosmic X-ray background or instrumental background. We fit the spectrum from 0.8 to 5 keV.

\begin{figure*}
\includegraphics[scale=0.45]{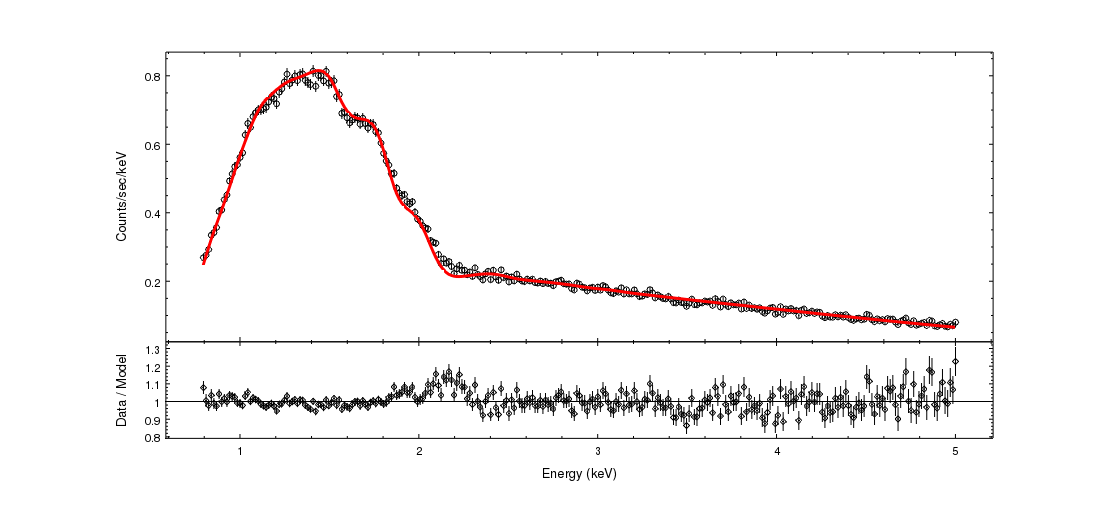}
\caption{The full fit shown between 0.8 and 5 keV using an \emph{xswabs} $\ti $ (\emph{powlaw1d + xsbapec})
model. The reduced $\chi^2$ is 1.45, a consequence of the excess around 2 keV.}
\label{FullFit}
\end{figure*}

\begin{figure*}
\includegraphics[scale=0.55]{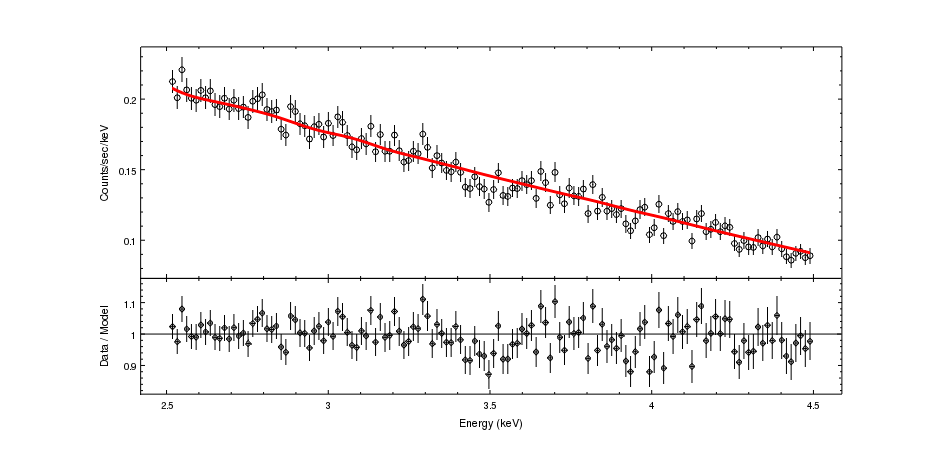}
\caption{The spectrum around 3.5 keV prior to the inclusion of a negative Gaussian. The fit is from 0.8 to 5 keV using a \emph{xswabs} $\ti $ (\emph{powlaw1d + xsbapec})
model, excluding the region from 1.8 to 2.3 keV containing a sharp effective area dip.}
\label{NoGaussian}
\end{figure*}

 The resulting fit is shown in Figure \ref{FullFit}.
This fit shows two main features, an excess around 2 - 2.2 keV and a deficit around 3.4 - 3.5 keV. The former is around the location of a sharp effective area dip in the telescope, which complicates its interpretation. Excesses around the effective area dip can arise from piled up Chandra observations, and so we do not discuss the feature around 2 keV any further. Our focus here is instead on the deficit around 3.4 - 3.5 keV, which is a clean part of the spectrum. We therefore refit excluding
the 1.8 - 2.3 keV region. This overall fit is now good, with a $\chi^2$ of 273 for 250 degrees of freedom. We show this fit, restricted to between
2.5 and 4.5 keV, in Figure \ref{NoGaussian}.

This fit clearly shows a deficit around 3.5 keV. To characterise the location and significance of this deficit precisely,
we then include a negative Gaussian in the fit, \emph{xswabs} $\ti $ (\emph{powlaw1d + xsbapec - xszgauss}). Here \emph{xszgauss} refers
to a Gaussian lineshape redshifted by $z$ from its rest frame value.
We allow the energy of the Gaussian to float between a range of 3.3 and 3.7 keV in the frame of the Perseus cluster (the observed energy is
redshifted by $z=0.0176$). For definiteness we assign the Gaussian lineshape an intrinsic width of 10 eV, but in practice any intrinsic width significantly smaller than the instrumental broadening of $\sim 100$ {\rm eV} is equivalent from a fitting perspective.

\begin{figure*}
\includegraphics[scale=0.55]{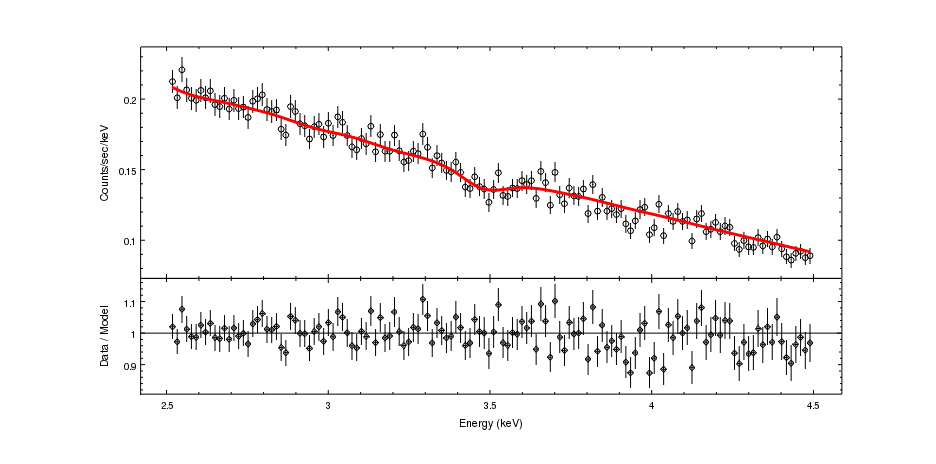}
\caption{The spectrum around 3.5 keV, now fitting the same region with \emph{xswabs} $\ti $ (\emph{powlaw1d + xsbapec - xszgauss}). The redshift of the Gaussian is fixed to $z=0.0176$
and its intrinsic line width fixed to $10 \, {\rm eV}$, with the norm and energy allowed to float.}
\label{WithGaussian}
\end{figure*}

The resulting fit is plotted in Figure \ref{WithGaussian} and we
see that the additional negative Gaussian is an excellent model for the deficit.
The best fit energy is $E = (3.54 \pm 0.02) \, \rm{keV}$, with an improvement of $\Delta \chi^2 = 20.0$ for an additional 2 degrees of freedom.
The best fit strength for the deficit is $( -7.8 \pm 1.7)$ photons ${\rm cm}^{-2}$ ${\rm s}^{-2}$.

\begin{figure}
\includegraphics[scale=0.65]{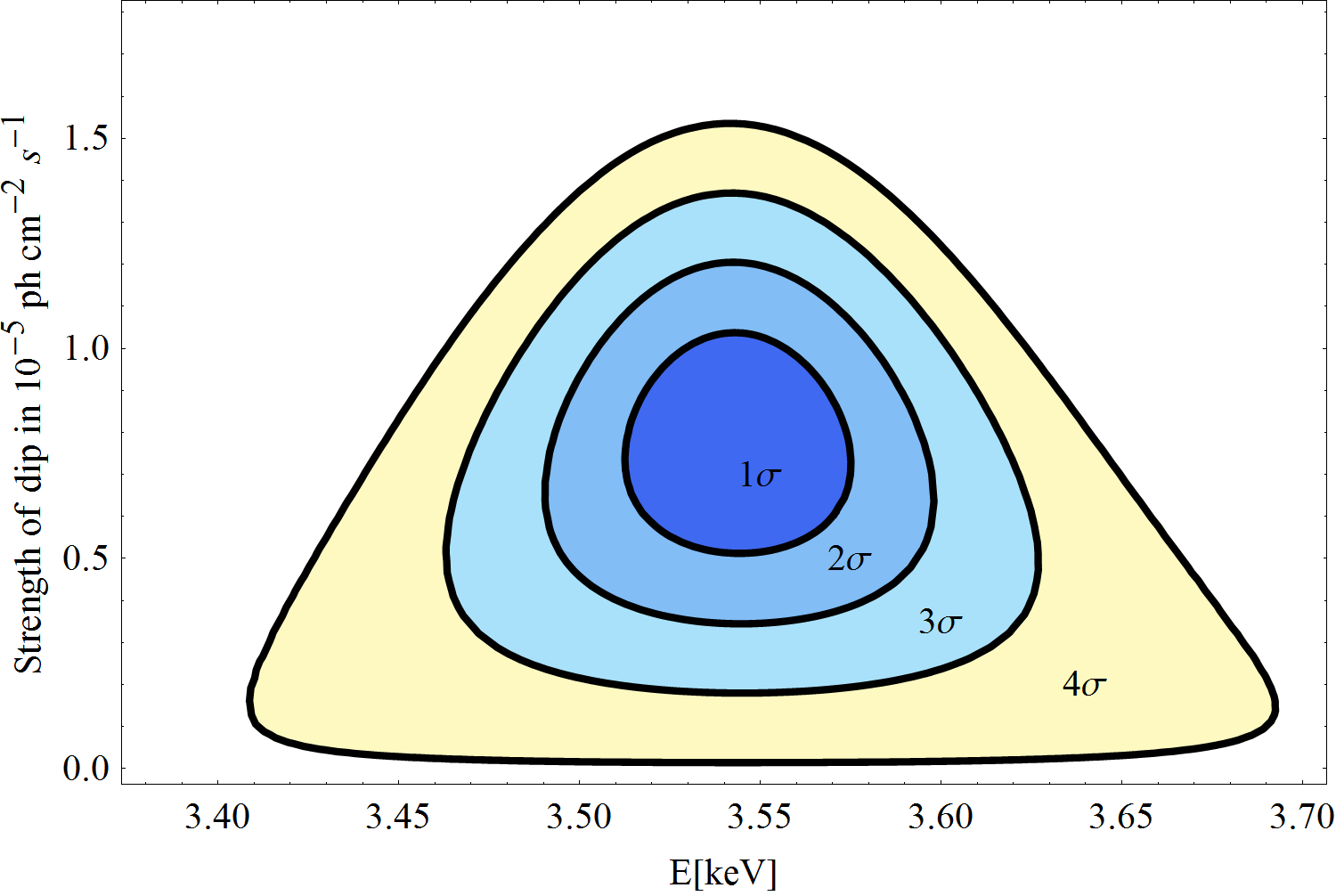}
\caption{The location and strength of the best-fit dip in the AGN spectrum, derived from stacked \emph{Chandra}
observations 11713, 12025, 12033 and 12036. \label{fig:contour}}
\end{figure}

For this fit, we show in Figure~\ref{fig:contour} a contour plot of the energy and strength of the negative
Gaussian.
As a check on the significance implied by the $\Delta \chi^2$ value, we generated 20,000 fake spectra~\footnote{At 4-sigma one in 16,000 spectra is an outlier, i.e. a 4-sigma dip would occur statistically at this rate.}
using Sherpa's \emph{fakepha} command (which produces samples of fake data for a given model) and the response files for the off-centre observations. We fit the fake data sets over the same energy
range and in the same manner as for the real data, first with
\emph{xswabs} $\ti $ (\emph{powlaw1d + xsbapec}) and then with \emph{xswabs} $\ti $ (\emph{powlaw1d + xsbapec - xszgauss}), allowing the energy of the Gaussian
 to float between 3.3 and 3.7 keV. However, for none of these fake data sets
was an additional negative Gaussian able to produce an improvement in the $\chi^2$ greater than that seen in the real data.

The above analysis focussed on the four \emph{Chandra} observations in which the AGN is highly off-axis.
As a side comment, we note that
there were also four further observations taken in 2009 (\emph{Chandra} obsids 11714, 11715, 11716, 12037) where the AGN is a
few arcminutes off-axis. Pileup remains a serious contaminant for these observations.
Nonetheless, we can exclude the central piled-up parts of the AGN image
to extract a `clean' spectrum.

We fit this in an identical fashion, working from 0.8 to 5 keV excluding 1.8 to 2.3 keV with a \emph{xswabs} $\ti $ (\emph{powlaw1d + xsbapec - xszgauss}) model.
Doing so, there is a mild
preference ($\Delta \chi^2 = 1.5$) for a dip in the AGN spectrum at a best-fit energy of $(3.55 \pm 0.10)$ keV.
If the overall fitted amplitude of the AGN spectrum in these observations is rescaled to be the same as that for the highly off-centre observations, the
amplitude of this dip would then correspond to $(-5.1 \pm 4.2) \ti 10^{-6} \, {\rm ph} \, {\rm cm}^{-2} \, {\rm s}^{-1}$. While insignificant by itself, this is consistent with the
earlier result using the highly off-axis observations.

It is striking that the dip in the AGN spectrum occurs at
an identical energy to the excess from the diffuse cluster emission.
We discuss possible physical models for this below.
First though, we note that
if such a dip is present in the AGN spectrum -- and it is over 4$\sigma$ significant in the only
clean observations of the AGN --
then it contributes to the \emph{Hitomi} data, which sums the AGN and cluster emission.

To determine the magnitude of its effect there,
we also require the overall AGN normalisation, as
on physical grounds such a dip only makes sense as a
fractional reduction in the AGN spectrum.
The AGN luminosity is highly variable; its lightcurve since 1970 is described in \cite{Fabian:2015kua}. Although still much dimmer than its 1980s peak, it
has been brightening since 2001. The clean 2009 \emph{Chandra} observations give a best-fit
normalisation of $4.7 \ti 10^{-3} \, {\rm ph} \, {\rm cm}^{-2} \, {\rm s}^{-1} \, {\rm keV}^{-1}$
at 1 keV, while it is reported in \cite{Aharonian:2016gzq} that the 2016 \emph{Hitomi} data give a normalisation
of $9.0 \ti 10^{-3} \, {\rm ph} \, {\rm cm}^{-2} \, {\rm s}^{-1} \, {\rm keV}^{-1}$ at 1 keV, an approximate doubling of the strength since 2009.

Based on this, we now rescale the 2009 result into an expected 2016 dip of $(-14.9 \pm 3.3) \ti 10^{-6} \, {\rm ph} \, {\rm cm}^{-2} \, {\rm s}^{-1}$ in
the AGN spectrum at $E = (3.54 \pm 0.02)$ keV. In \cite{Aharonian:2016gzq} it is reported that the expected diffuse
excess emission in the 2016 \emph{Hitomi} data, based on \emph{XMM-Newton} observations restricted to the SXS field of view, was $(9.0 \pm 2.9) \ti
10^{-6} \, {\rm ph} \, {\rm cm}^{-2} \, {\rm s}^{-1}$ at $E = 3.54_{-0.04}^{+0.03}$ keV.
Summing these two results then leads us to
expect a \emph{dip} in the 2016 \emph{Hitomi} data at 3.54 keV of $(-5.9 \pm 4.4) \ti 10^{-6} \, {\rm ph} \, {\rm cm}^{-2} \, {\rm s}^{-1}$.

And indeed this is precisely what is observed: from Figure 3 in \cite{Aharonian:2016gzq}, we see that for broadening by the dark matter velocity dispersion
of $1300 \, {\rm km} \, {\rm s}^{-1}$ the \emph{Hitomi} data shows a
best-fit dip of $(-8 \ti 10^{-6}) \, {\rm ph} \, {\rm cm}^{-2}
\, {\rm s}^{-1}$ at $E = (3.55 \pm 0.02)$ keV (the error is estimated from Figure 3 in \cite{Aharonian:2016gzq}), at
 an approximate significance of $2.5 \sigma$ ($\Delta \chi^2 \sim 7$).

The data sets are therefore consistent; \emph{XMM-Newton} data shows an excess in the diffuse cluster emission at $3.54^{+ 0.03}_{-0.04} {\rm keV}$,
\emph{Chandra} data on the AGN shows a strong dip in the spectrum at the same energy, and
\emph{Hitomi} data (sensitive to both cluster and AGN) gives a dip
at $(3.55 \pm 0.02) {\rm  keV}$ of the expected magnitude once these two effects are combined.

\subsection{Fluorescent Dark Matter}

In the previous section, we saw that the
various datasets on 3.5 keV photons from Perseus are remarkably consistent and contain
a variety of high-significance features at $E \simeq 3.54~{\rm keV}$.

We now consider possible interpretations. We focus on ways to obtain deficits in the AGN spectrum; there are many
ways, both astrophysical and exotic, to generate
excesses, but fewer that can give rise to spectral deficits.
Instrumental explanations are also less plausible here. The equivalent width of the dip (this is defined as 
the width of the continuum emission that would give the same number of photons that are absent due to the dip, and so is a measure
of the strength of a line relative to the surrounding continuum)
is $\sim 15 {\rm eV}$. This compares to a (much weaker) equivalent width of $\sim 1 {\rm eV}$ for the diffuse excess reported in 
\cite{Bulbul:2014sua}.

One way to obtain a deficit in an AGN spectrum is via axion-like particles \cite{Wouters:2013hua, Berg:2016ese}, as photons can convert to
 axion-like particles in the presence of astrophysical magnetic fields. While axion-like particles have been proposed as models to explain the morphology of the
 3.5 keV diffuse excess \cite{Cicoli:2014bfa}, it would require a considerable coincidence to produce a single dip in the AGN spectrum at an identical
 energy to the diffuse cluster excess.

We therefore look for models where the same underlying mechanism is responsible for both the deficit in the
AGN spectrum and the excess in the diffuse spectrum at 3.54 keV (3.48 keV in the observer frame).
A simple scenario involves a 2-state dark matter model ($\chi_1$ and $\chi_2$). The lower state $\chi_1$ absorbs a 3.54 keV photon
to enter the excited state $\chi_2$, which then decays by re-emission of the photon. A sample Lagrangian is
\begin{equation}
\mathcal{L} \supset \frac{1}{M} \bar{\chi}_{2} \sigma_{\mu \nu}  \chi_1 F^{\mu \nu},
\end{equation}
and this type of resonant absorption is analysed in greater detail in \cite{Profumo:2006im}.
The AGN dip at $E = 3.54~{\rm keV}$ would then be
a dark sector analogue of e.g.~Lyman-$\alpha$ absorption, with the diffuse excess at the same energy the result of fluorescent re-emission.

To explain both the 3.5 keV deficit and excess in this way, the states $\chi_1$ and $\chi_2$ must be massive enough that they can be treated as
non-relativistic -- otherwise, the line arising from the decay $\chi_2 \to \chi_1 \gamma$ would be Doppler shifted, and so appear at a different energy
than for the absorption process $\chi_1 + \gamma \to \chi_2$. As the energy of the dip $(3.54 \pm 0.02) {\rm keV}$ is consistent at the per cent level
with the 3.5 keV emission line, this requires $m_1 > {\rm 1 MeV}$. For the rest of this paper we shall assume that this holds, so that the
particles $\chi_1$ and $\chi_2$ can be treated for the purposes of emission and absorption as extremely heavy and non-relativistic.

If we treat the absorption of 3.54 keV photons as a Breit-Wigner resonance and assume a 100\% branching ratio for
$\chi_2 \to \chi_1 \gamma$, we can determine the width $\Gamma$ of the process $\chi_2 \to \chi_1 \gamma$ using the
observed photon deficit from the AGN.
The cross section for $\chi_1 \gamma \to \chi_1 \gamma$ has a resonance at photon energy $E_0 = \frac{m_2^2 - m_1^2}{2 m_1} \stackrel{!}{=} 3.54 \, {\rm keV}$ in the rest frame of the dark matter, where $m_1$ and $m_2$ are the masses of $\chi_1$ and $\chi_2$ respectively. Near the resonance, the
cross section is described by the relativistic Breit-Wigner formula:
\begin{equation}
\sigma_{\rm BW} (E) = \frac{2 \pi} {p_{CM}^2} \frac{\Gamma_{\chi_2 \to \chi_1 \gamma}}{\Gamma_{\chi_2}} \frac{(m_2 \Gamma_{\chi_2})^2}{(s - m_2^2)^2 + (m_2 \Gamma_{\chi_2})^2}~,
\end{equation}
where $p^2_{CM} = \frac{m_1^2 E^2}{m_1^2 + 2 m_1 E}$ is the squared magnitude of the momentum in the centre of mass frame; $\Gamma_{\chi_2 \to \chi_1 \gamma}$ is the decay rate of $\chi_2$ to $\chi_1 \gamma$; $\Gamma_{\chi_2}$ is the total decay rate of $\chi_2$ and $\sqrt{s}$ is the centre of mass energy. As we assume a 100\% $\chi_2 \to \chi_1 + \gamma$ branching ratio, we have $\Gamma{\chi_2 \to \chi_1 \gamma} = \Gamma_{\chi_2} = \Gamma$, and we shall also write $m_{DM} \equiv m_1 \simeq m_2$, as the mass difference between the two states is much smaller than their absolute values.

For the dark matter column density along the line of sight to NGC1275, we use an NFW profile appropriate to the Perseus cluster
\begin{equation}
\rho_{DM}(r) = \frac{\rho_0}{\frac{r}{r_s} \left( 1 + \left( \frac{r}{r_s} \right)^2 \right)}~,
\end{equation}
with $r_s = 0.477$ Mpc and $\rho_0 = 7.35 \times 10^{14} M_{\odot} \, {\rm Mpc}^{3}$ \cite{1104.3530}.
As the AGN is the dynamical centre of the cluster, the integrated column density is formally divergent, and we cut off the integral at
0.01 and 2 Mpc.

However, there is a significant uncertainty attached to the exact column density towards NGC1275. There are two main factors.

First, the actual Perseus cluster profile may be cored instead of the cusp
present in NFW. For low-mass galaxies, the profile is known to have a core rather than a cusp. This may arise merely from
baryonic feedback on a cuspy cold dark matter profile, but alternatively may requires a more substantial modification such as warm dark matter.
If the Perseus cluster profile is indeed cored, it will reduce the column density from the value inferred above.

Secondly, the profile is only that for the cluster. The line of sight originates at the very centre of NGC1275, the supergiant central
elliptical galaxy of the cluster. The NGC1275 dark matter profile is not known but is expected to give a significant contribution to the
dark matter column density (for example, given the size of NGC1275
it should be much larger than that towards Sgr A*).

Given that the dip strength
has equivalent width of 15 eV, we can then derive (assuming $m_{DM} \gg 3.5 {\rm keV}$)
\begin{equation}
\Gamma \gtrsim \left( \frac{m_{DM}}{{\rm GeV}} \right) \ti (1 - 10) \ti 10^{-10} \, {\rm keV}~.
\label{GammaEq}
\end{equation}
The steps to obtain this result are as follows. First, the equivalent width of 15eV implies (by definition) that, from a flat distribution of
photons between (for definiteness) 3.25 and 3.85 keV, 2.5\% of them must be absorbed, thereby generating the dip. Second, the photon opacity $\tau(E)$ at any give energy $E$ is found by taking the product of the dark matter column density along the line of sight and the absorption (Breit-Wigner) cross-section
convolved with the dark matter velocity dispersion. The equivalent width (EW) is then given by
\begin{equation}
EW = 15 {\rm eV} = \int 1 - \exp{(-\tau(E))} \, dE,
\label{opacity}
\end{equation}
with lower (upper) limits well below (above) the central resonant energy. As $\Gamma$ enters the Breit-Wigner cross-section, 
equation (\ref{opacity}) can be solved numerically for $\Gamma$. Note that the shape of the dip (i.e. the opacity) is determined dominantly by the
velocity broadening rather than the intrinsic width.

We give a broad range to allow for the significant uncertainty in the actual dark matter
column density along the line of sight to NGC1275. The inequality in equation (\ref{GammaEq}) 
arises because the equivalent width of the dip strength is very similar to that expected to be induced by
dark matter broadening ($\sigma_{DM, velocity, Perseus} \sim 15 {\rm eV}$ \cite{Aharonian:2016gzq}). As for any energy absorption can never exceed 100\%, once absorption becomes saturated
within a region, further increases in the
equivalent width of the dip only increase logarithmically with $\Gamma$, as it must come from the tail of the velocity-broadening Gaussian.

We also note that such a dip could not arise from absorption onto an atomic line. Besides the fact that no absorption lines
are known around $E \sim 3.55 {\rm keV}$, the thermal velocity broadening in Perseus of $\sigma_v \sim 3 \, {\rm eV}$ \cite{Aharonian:2016gzq}
is far narrower than the measured equivalent width. 

\subsection{Morphological Features}

We now consider the morphological distribution of the 3.5 keV emission. We consider the simplest case, where all 3.5 keV emission in
$\chi_2 \to \chi_1 \gamma$ arises after initial absorption of a real 3.5 keV photon. More generally, one could also
consider cases where the absorbed photon is virtual and arises from scattering off protons, electrons~\cite{D'Eramo:2016xxw} or other particles.

For this simplest case, the most basic feature of fluorescent dark matter is that the total number of 3.5 keV photons is conserved: the total excess
emission, integrated across a cluster, must be precisely balanced by the integrated deficit. This result is independent of the detailed dark matter profile;
as all absorbed photons are subsequently re-emitted, it follows that there is no net production of 3.5 keV photons in this model.

Applying this to the Perseus cluster, this would require the time-averaged \emph{deficit} in 3.5 keV photons from the central AGN,
measured in units of ${\rm ph} \, {\rm cm}^{-2} \, {\rm s}^{-1}$, to
precisely equal the \emph{excess} in 3.5 keV photons from diffuse emission across the entire cluster. The
2009 \emph{Chandra} dip is $-7.8 \ti 10^{-6} \, {\rm ph} \, {\rm cm}^{-2} \, {\rm s}^{-1}$, while the total excess reported
in \cite{Bulbul:2014sua} across the \emph{XMM-Newton} field of view is $52^{+24}_{-15} \ti 10^{-6} \, {\rm ph} \, {\rm cm}^{-2} \, {\rm s}^{-1}$ (given the large
field of view of \emph{XMM-Newton}, we take this value as a proxy for the total emission of the cluster).

For a time-varying AGN luminosity, the photon density at radius $r$ depends on the luminosity at time $(t_{now} - \frac{r}{c})$.
The observations of the 3.5 keV excess are based on fields of view extending over
$\mc{O}(10 - 100 {\rm kpc})$ regions. In this scenario, the magnitude of the excess then requires effectively averaging
the AGN luminosity over periods of $10^4 - 10^6$ years.
Given our (limited) knowledge of the time-variability of the AGN, these
values are consistent. For example, from 1970 - 1988 the AGN luminosity was 5-8 times greater than its 2009 value \cite{Fabian:2015kua}, and if this period represents a typical
luminosity over the last $10^6$ years then the observed deficits and excesses are consistent.

In fluorescent dark matter, the magnitude of the diffuse emission is set by the product of the dark matter density and the photon density.
We first consider only the photons originating from the AGN and
approximate the AGN emission as spherically symmetric and with constant luminosity. In this case
$\rho_{\gamma}(r, \theta, \phi, t) \propto r^{-2}$. For an NFW profile,
the 3.5 keV emissivity is then
$$
\mc{L}_{3.5 {\rm keV}} \propto \rho_{DM}(r) \rho_{\gamma}(r) \propto \frac{\rho_0}{r^3 \left( 1 + \left( \frac{r}{r_s} \right) \right)^2}~.
$$
This has a much sharper central peaking than either decaying ($\propto \rho_{DM}(r)$) or annihilating dark matter ($\propto\rho_{DM}(r)^2$).
This is interesting as a sharp central peaking is preferred by the results of \cite{Bulbul:2014sua, Urban:2014yda, Franse:2016dln} (indeed \cite{Bulbul:2014sua}
found that approximately half
the 3.5 keV diffuse emission was contained within $1'$ of the cluster centre).

\begin{figure}
\includegraphics[width=0.45\textwidth]{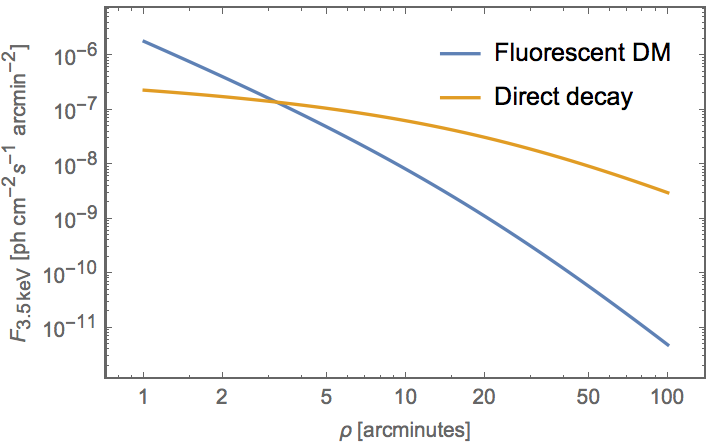}
\caption{The radial profile of the 3.5 keV line flux for the fluorescent dark matter model presented here and for dark matter decay. The flux is normalised
to match the overall flux in \cite{Bulbul:2014sua}.} \label{RadialProfile}
\end{figure}

Using an NFW dark matter profile for simplicity, we plot in Figure~\ref{RadialProfile} the resulting radial emission profile.
For comparison, we also plot the radial profile of a 3.5 keV line from direct dark matter decay to photons normalised
to give the same total flux integrated over the cluster. In the direct decay case, the flux is proportional to the dark matter column density.

This result will be modified as a consequence of absorption and re-emission of the diffuse thermal 3.5 keV photons present from the thermal bremsstrahlung spectrum.
As the overall number of photons is conserved, this will not affect the overall magnitude of the diffuse excess but
will modify the radial profile. Another modification of the radial profile could arise if the dark matter profile is cored rather than cuspy as for NFW; in this case the
central peaking of the excess would be $\propto r^{-2}$ rather than $\propto r^{-3}$.

A further modification would be due to the angular anisotropy of the AGN flux. The jets have preferred orientations, and in this scenario we would
expect the excess to be preferentially generated near to the axes of the jets. Given high-statistics evidence for the diffuse 3.5 keV excess, one could search
for an angular anisotropy in the flux correlated with the direction of the jet. However, as currently there is only approximately 4$\sigma$ evidence for the line
from the entirety of the Perseus cluster, there are insufficient statistics to do a detailed morphological analysis.
We therefore defer a detailed study of angular anisotropies in this model to future work.

In the fluorescent dark matter scenario, there is no net production of 3.5 keV photons.
Note that this does not lead to a contradiction with the observation of a net excess from the stacked cluster sample of \cite{Bulbul:2014sua}.
A general feature of this model is that it leads to deficits in 3.5 keV photons from
point sources and excesses from the diffuse emission; in the stacked analysis, however, all point sources are removed from the
analysis \emph{ab initio} (using \emph{wavdetect}) and the spectrum only contains the diffuse emission. A possible signal of this model could be found by observing
a deficit at 3.54 keV in stacked analyses of point sources, weighted by the line-of-sight dark matter column density.

\subsection{Conclusions}

We have argued that \emph{Hitomi}, \emph{XMM-Newton} and \emph{Chandra} observations of the Perseus cluster at $E \sim 3.5 {\rm keV}$
show a remarkable degree of consistency. In particular, the Hitomi spectrum around 3.5 keV can be understood as the sum of a dip in the
AGN spectrum at $E = (3.54 \pm 0.02)$ keV (observed by \emph{Chandra}) with an excess in the diffuse
cluster emission at an identical energy (observed by \emph{XMM-Newton} and \emph{Chandra}).
We have described dark matter models that can give rise to this phenomenology.

Sadly \emph{Hitomi} is no longer able to contribute to observational efforts to understand the 3.5 keV line.
We have emphasised that an accurate and clean spectrum of the NGC1275 AGN is crucial for understanding this phenomenon.
Significant improvements on this can be made using operating satellites and with existing CCD technology. The best current
spectrum was taken in 2009 by $\emph{Chandra}$ with the nominal frame time of 3 seconds.
Given the AGN is now twice as bright, a further dedicated off-axis observation of NGC1275, operating with reduced
frame time to minimise pileup, would give a substantial improvement over the 2009 data.

\subsection{Acknowledgments}
This project is funded in part by the European Research Council starting grant `Supersymmetry Breaking in String Theory' (307605).
JC is also funded by a Royal Society University Research Fellowship. We thank Marcus Berg and the referees for comments on the paper.

\bibliography{hitomi35refs} 

\begin{thebibliography}{19}
\expandafter\ifx\csname natexlab\endcsname\relax\def\natexlab#1{#1}\fi
\expandafter\ifx\csname bibnamefont\endcsname\relax
  \def\bibnamefont#1{#1}\fi
\expandafter\ifx\csname bibfnamefont\endcsname\relax
  \def\bibfnamefont#1{#1}\fi
\expandafter\ifx\csname citenamefont\endcsname\relax
  \def\citenamefont#1{#1}\fi
\expandafter\ifx\csname url\endcsname\relax
  \def\url#1{\texttt{#1}}\fi
\expandafter\ifx\csname urlprefix\endcsname\relax\def\urlprefix{URL }\fi
\providecommand{\bibinfo}[2]{#2}
\providecommand{\eprint}[2][]{\url{#2}}

\bibitem[{\citenamefont{Bulbul et~al.}(2014)\citenamefont{Bulbul, Markevitch,
  Foster, Smith, Loewenstein, and Randall}}]{Bulbul:2014sua}
\bibinfo{author}{\bibfnamefont{E.}~\bibnamefont{Bulbul}},
  \bibinfo{author}{\bibfnamefont{M.}~\bibnamefont{Markevitch}},
  \bibinfo{author}{\bibfnamefont{A.}~\bibnamefont{Foster}},
  \bibinfo{author}{\bibfnamefont{R.~K.} \bibnamefont{Smith}},
  \bibinfo{author}{\bibfnamefont{M.}~\bibnamefont{Loewenstein}},
  \bibnamefont{and} \bibinfo{author}{\bibfnamefont{S.~W.}
  \bibnamefont{Randall}}, \bibinfo{journal}{Astrophys. J.}
  \textbf{\bibinfo{volume}{789}}, \bibinfo{pages}{13} (\bibinfo{year}{2014}),
  \eprint{1402.2301}.

\bibitem[{\citenamefont{Boyarsky et~al.}(2014)\citenamefont{Boyarsky,
  Ruchayskiy, Iakubovskyi, and Franse}}]{Boyarsky:2014jta}
\bibinfo{author}{\bibfnamefont{A.}~\bibnamefont{Boyarsky}},
  \bibinfo{author}{\bibfnamefont{O.}~\bibnamefont{Ruchayskiy}},
  \bibinfo{author}{\bibfnamefont{D.}~\bibnamefont{Iakubovskyi}},
  \bibnamefont{and} \bibinfo{author}{\bibfnamefont{J.}~\bibnamefont{Franse}},
  \bibinfo{journal}{Phys. Rev. Lett.} \textbf{\bibinfo{volume}{113}},
  \bibinfo{pages}{251301} (\bibinfo{year}{2014}), \eprint{1402.4119}.

\bibitem[{\citenamefont{Carlson et~al.}(2015)\citenamefont{Carlson, Jeltema,
  and Profumo}}]{1411.1758}
\bibinfo{author}{\bibfnamefont{E.}~\bibnamefont{Carlson}},
  \bibinfo{author}{\bibfnamefont{T.}~\bibnamefont{Jeltema}}, \bibnamefont{and}
  \bibinfo{author}{\bibfnamefont{S.}~\bibnamefont{Profumo}},
  \bibinfo{journal}{JCAP} \textbf{\bibinfo{volume}{1502}}, \bibinfo{pages}{009}
  (\bibinfo{year}{2015}), \eprint{1411.1758}.

\bibitem[{\citenamefont{Urban et~al.}(2015)\citenamefont{Urban, Werner, Allen,
  Simionescu, Kaastra, and Strigari}}]{Urban:2014yda}
\bibinfo{author}{\bibfnamefont{O.}~\bibnamefont{Urban}},
  \bibinfo{author}{\bibfnamefont{N.}~\bibnamefont{Werner}},
  \bibinfo{author}{\bibfnamefont{S.~W.} \bibnamefont{Allen}},
  \bibinfo{author}{\bibfnamefont{A.}~\bibnamefont{Simionescu}},
  \bibinfo{author}{\bibfnamefont{J.~S.} \bibnamefont{Kaastra}},
  \bibnamefont{and} \bibinfo{author}{\bibfnamefont{L.~E.}
  \bibnamefont{Strigari}}, \bibinfo{journal}{Mon. Not. Roy. Astron. Soc.}
  \textbf{\bibinfo{volume}{451}}, \bibinfo{pages}{2447} (\bibinfo{year}{2015}),
  \eprint{1411.0050}.

\bibitem[{\citenamefont{Franse et~al.}(2016)}]{Franse:2016dln}
\bibinfo{author}{\bibfnamefont{J.}~\bibnamefont{Franse}} \bibnamefont{et~al.}
  (\bibinfo{year}{2016}), \eprint{1604.01759}.

\bibitem[{\citenamefont{Tamura et~al.}(2015)\citenamefont{Tamura, Iizuka,
  Maeda, Mitsuda, and Yamasaki}}]{Tamura:2014mta}
\bibinfo{author}{\bibfnamefont{T.}~\bibnamefont{Tamura}},
  \bibinfo{author}{\bibfnamefont{R.}~\bibnamefont{Iizuka}},
  \bibinfo{author}{\bibfnamefont{Y.}~\bibnamefont{Maeda}},
  \bibinfo{author}{\bibfnamefont{K.}~\bibnamefont{Mitsuda}}, \bibnamefont{and}
  \bibinfo{author}{\bibfnamefont{N.~Y.} \bibnamefont{Yamasaki}},
  \bibinfo{journal}{Publ. Astron. Soc. Jap.} \textbf{\bibinfo{volume}{67}},
  \bibinfo{pages}{23} (\bibinfo{year}{2015}), \eprint{1412.1869}.

\bibitem[{\citenamefont{Gu et~al.}(2015)\citenamefont{Gu, Kaastra, Raassen,
  Mullen, Cumbee, Lyons, and Stancil}}]{GuKaastra}
\bibinfo{author}{\bibfnamefont{L.}~\bibnamefont{Gu}},
  \bibinfo{author}{\bibfnamefont{J.}~\bibnamefont{Kaastra}},
  \bibinfo{author}{\bibfnamefont{A.~J.~J.} \bibnamefont{Raassen}},
  \bibinfo{author}{\bibfnamefont{P.~D.} \bibnamefont{Mullen}},
  \bibinfo{author}{\bibfnamefont{R.~S.} \bibnamefont{Cumbee}},
  \bibinfo{author}{\bibfnamefont{D.}~\bibnamefont{Lyons}}, \bibnamefont{and}
  \bibinfo{author}{\bibfnamefont{P.~C.} \bibnamefont{Stancil}},
  \bibinfo{journal}{Astron. Astrophys.} \textbf{\bibinfo{volume}{584}},
  \bibinfo{pages}{L11} (\bibinfo{year}{2015}), \eprint{1511.06557}.

\bibitem[{\citenamefont{Iakubovskyi}(2015)}]{151000358}
\bibinfo{author}{\bibfnamefont{D.}~\bibnamefont{Iakubovskyi}}
  (\bibinfo{year}{2015}), \eprint{1510.00358}.

\bibitem[{\citenamefont{Aharonian et~al.}(2017)}]{Aharonian:2016gzq}
\bibinfo{author}{\bibfnamefont{F.~A.} \bibnamefont{Aharonian}}
  \bibnamefont{et~al.} (\bibinfo{collaboration}{Hitomi}),
  \bibinfo{journal}{Astrophys. J.} \textbf{\bibinfo{volume}{837}},
  \bibinfo{pages}{L15} (\bibinfo{year}{2017}), \eprint{1607.07420}.

\bibitem[{\citenamefont{Berg et~al.}(2016)\citenamefont{Berg, Conlon, Day,
  Jennings, Krippendorf, Powell, and Rummel}}]{Berg:2016ese}
\bibinfo{author}{\bibfnamefont{M.}~\bibnamefont{Berg}},
  \bibinfo{author}{\bibfnamefont{J.~P.} \bibnamefont{Conlon}},
  \bibinfo{author}{\bibfnamefont{F.}~\bibnamefont{Day}},
  \bibinfo{author}{\bibfnamefont{N.}~\bibnamefont{Jennings}},
  \bibinfo{author}{\bibfnamefont{S.}~\bibnamefont{Krippendorf}},
  \bibinfo{author}{\bibfnamefont{A.~J.} \bibnamefont{Powell}},
  \bibnamefont{and} \bibinfo{author}{\bibfnamefont{M.}~\bibnamefont{Rummel}},
  \bibinfo{journal}{Astrophys. Journal (in press).}  (\bibinfo{year}{2016}),
  \eprint{1605.01043}.

\bibitem[{\citenamefont{Aharonian et~al.}(2016)}]{Aharonian:2016pyf}
\bibinfo{author}{\bibfnamefont{F.}~\bibnamefont{Aharonian}}
  \bibnamefont{et~al.} (\bibinfo{collaboration}{Hitomi})
  (\bibinfo{year}{2016}), \eprint{1607.04487}.

\bibitem[{\citenamefont{{Fruscione} et~al.}(2006)\citenamefont{{Fruscione},
  {McDowell}, {Allen}, {Brickhouse}, {Burke}, {Davis}, {Durham}, {Elvis},
  {Galle}, {Harris} et~al.}}]{ciao}
\bibinfo{author}{\bibfnamefont{A.}~\bibnamefont{{Fruscione}}},
  \bibinfo{author}{\bibfnamefont{J.~C.} \bibnamefont{{McDowell}}},
  \bibinfo{author}{\bibfnamefont{G.~E.} \bibnamefont{{Allen}}},
  \bibinfo{author}{\bibfnamefont{N.~S.} \bibnamefont{{Brickhouse}}},
  \bibinfo{author}{\bibfnamefont{D.~J.} \bibnamefont{{Burke}}},
  \bibinfo{author}{\bibfnamefont{J.~E.} \bibnamefont{{Davis}}},
  \bibinfo{author}{\bibfnamefont{N.}~\bibnamefont{{Durham}}},
  \bibinfo{author}{\bibfnamefont{M.}~\bibnamefont{{Elvis}}},
  \bibinfo{author}{\bibfnamefont{E.~C.} \bibnamefont{{Galle}}},
  \bibinfo{author}{\bibfnamefont{D.~E.} \bibnamefont{{Harris}}},
  \bibnamefont{et~al.}, in \emph{\bibinfo{booktitle}{Society of Photo-Optical
  Instrumentation Engineers (SPIE) Conference Series}} (\bibinfo{year}{2006}),
  vol. \bibinfo{volume}{6270} of \emph{\bibinfo{series}{Proceedings of the
  International Society for Optical Engineering}}, p. \bibinfo{pages}{62701V}.

\bibitem[{\citenamefont{{Freeman} et~al.}(2001)\citenamefont{{Freeman}, {Doe},
  and {Siemiginowska}}}]{sherpa}
\bibinfo{author}{\bibfnamefont{P.}~\bibnamefont{{Freeman}}},
  \bibinfo{author}{\bibfnamefont{S.}~\bibnamefont{{Doe}}}, \bibnamefont{and}
  \bibinfo{author}{\bibfnamefont{A.}~\bibnamefont{{Siemiginowska}}}, in
  \emph{\bibinfo{booktitle}{Astronomical Data Analysis}}, edited by
  \bibinfo{editor}{\bibfnamefont{J.-L.} \bibnamefont{{Starck}}}
  \bibnamefont{and} \bibinfo{editor}{\bibfnamefont{F.~D.}
  \bibnamefont{{Murtagh}}} (\bibinfo{year}{2001}), vol. \bibinfo{volume}{4477}
  of \emph{\bibinfo{series}{Proceedings of the International Society for
  Optical Engineering}}, pp. \bibinfo{pages}{76--87},
  \eprint{astro-ph/0108426}.

\bibitem[{\citenamefont{Fabian et~al.}(2015)\citenamefont{Fabian, Walker,
  Pinto, Russell, and Edge}}]{Fabian:2015kua}
\bibinfo{author}{\bibfnamefont{A.~C.} \bibnamefont{Fabian}},
  \bibinfo{author}{\bibfnamefont{S.~A.} \bibnamefont{Walker}},
  \bibinfo{author}{\bibfnamefont{C.}~\bibnamefont{Pinto}},
  \bibinfo{author}{\bibfnamefont{H.~R.} \bibnamefont{Russell}},
  \bibnamefont{and} \bibinfo{author}{\bibfnamefont{A.~C.} \bibnamefont{Edge}},
  \bibinfo{journal}{Mon. Not. Roy. Astron. Soc.}
  \textbf{\bibinfo{volume}{451}}, \bibinfo{pages}{3061} (\bibinfo{year}{2015}),
  \eprint{1505.03754}.

\bibitem[{\citenamefont{Wouters and Brun}(2013)}]{Wouters:2013hua}
\bibinfo{author}{\bibfnamefont{D.}~\bibnamefont{Wouters}} \bibnamefont{and}
  \bibinfo{author}{\bibfnamefont{P.}~\bibnamefont{Brun}},
  \bibinfo{journal}{Astrophys. J.} \textbf{\bibinfo{volume}{772}},
  \bibinfo{pages}{44} (\bibinfo{year}{2013}), \eprint{1304.0989}.

\bibitem[{\citenamefont{Cicoli et~al.}(2014)\citenamefont{Cicoli, Conlon,
  Marsh, and Rummel}}]{Cicoli:2014bfa}
\bibinfo{author}{\bibfnamefont{M.}~\bibnamefont{Cicoli}},
  \bibinfo{author}{\bibfnamefont{J.~P.} \bibnamefont{Conlon}},
  \bibinfo{author}{\bibfnamefont{M.~C.~D.} \bibnamefont{Marsh}},
  \bibnamefont{and} \bibinfo{author}{\bibfnamefont{M.}~\bibnamefont{Rummel}},
  \bibinfo{journal}{Phys. Rev.} \textbf{\bibinfo{volume}{D90}},
  \bibinfo{pages}{023540} (\bibinfo{year}{2014}), \eprint{1403.2370}.

\bibitem[{\citenamefont{Profumo and Sigurdson}(2007)}]{Profumo:2006im}
\bibinfo{author}{\bibfnamefont{S.}~\bibnamefont{Profumo}} \bibnamefont{and}
  \bibinfo{author}{\bibfnamefont{K.}~\bibnamefont{Sigurdson}},
  \bibinfo{journal}{Phys. Rev.} \textbf{\bibinfo{volume}{D75}},
  \bibinfo{pages}{023521} (\bibinfo{year}{2007}), \eprint{astro-ph/0611129}.

\bibitem[{\citenamefont{Sanchez-Conde et~al.}(2011)\citenamefont{Sanchez-Conde,
  Cannoni, Zandanel, Gomez, and Prada}}]{1104.3530}
\bibinfo{author}{\bibfnamefont{M.~A.} \bibnamefont{Sanchez-Conde}},
  \bibinfo{author}{\bibfnamefont{M.}~\bibnamefont{Cannoni}},
  \bibinfo{author}{\bibfnamefont{F.}~\bibnamefont{Zandanel}},
  \bibinfo{author}{\bibfnamefont{M.~E.} \bibnamefont{Gomez}}, \bibnamefont{and}
  \bibinfo{author}{\bibfnamefont{F.}~\bibnamefont{Prada}},
  \bibinfo{journal}{JCAP} \textbf{\bibinfo{volume}{1112}}, \bibinfo{pages}{011}
  (\bibinfo{year}{2011}), \eprint{1104.3530}.

\bibitem[{\citenamefont{D'Eramo et~al.}(2016)\citenamefont{D'Eramo, Hambleton,
  Profumo, and Stefaniak}}]{D'Eramo:2016xxw}
\bibinfo{author}{\bibfnamefont{F.}~\bibnamefont{D'Eramo}},
  \bibinfo{author}{\bibfnamefont{K.}~\bibnamefont{Hambleton}},
  \bibinfo{author}{\bibfnamefont{S.}~\bibnamefont{Profumo}}, \bibnamefont{and}
  \bibinfo{author}{\bibfnamefont{T.}~\bibnamefont{Stefaniak}},
  \bibinfo{journal}{Phys. Rev.} \textbf{\bibinfo{volume}{D93}},
  \bibinfo{pages}{103011} (\bibinfo{year}{2016}), \eprint{1603.04859}.

\end{thebibliography}

\end{document}